# Microstructural characterization and kinetics modelling of vermicular cast irons.

Karina Laneri (♣), Pere Bruna(♦) and Daniel Crespo(♦).

(♣) Departamento de Física, Facultad de Ciencias Exactas, UNLP, IFLP-CONICET, CC67 (1900) La Plata, Argentina.

(♦) Departament de Física Aplicada, Escola Politècnica Superior de Castelldefels, and Centre de Recerca en Nanoenginyeria, Universitat Politècnica de Catalunya, Avda. del Canal Olímpic s/n , 08860-Castelldefels, Spain.

**Abstract**

Several experimental techniques are used for phase identification and microstructure characterization of austempered vermicular cast irons (XRD, SEM, TEM and Mössbauer spectroscopy). Acicular structures were found to be composed by ferrite and austenite with average sizes compatible with those reported for bainitic ferrite in steels and Austempered Ductile Iron. An assessment of the free energy change involved in the *austenite→bainite* transformation assuming a plate-like nucleation shape for bainite gave an average characteristic length close to the observed from statistical length distributions. The kinetics of the transformation was modelled in the Avrami framework; both the diffusion controlled and the diffusionless growth hypothesis were considered in order to elucidate the mechanism underlying the austempering phase transformation. Results indicated that diffusion of C is the responsible of the nucleation process of the bainite sheaves, that appear as a consequence of a localized displacive transformation when the C concentration is adequate, but further growth of the bainite plates is almost suppressed.

# INTRODUCTION

Cast irons contain higher C content than steels (more than 2 wt%), Si as a principal alloy component and other elements such as Mn that can be regulated to design a material with desirable mechanical properties. In particular, ductile or nodular cast iron contains trace amounts of magnesium which, by reacting with sulfur and oxygen in the molten iron, precipitates out carbon in the form of small spheres. These spheres improve the stiffness, strength, and shock resistance of ductile iron over gray iron. Accordingly, graphite vermicular morphology within cast irons is also obtained by addition of trace amounts of magnesium. Until recently, Cast irons with vermicular morphology, also referred to as Compacted Graphite (CG) Cast Irons have been extremely difficult to produce on a commercial scale because of process-control difficulties and the necessity of keeping alloy additions within very tight limits (if the Mg addition varied by as little as 0.005% results would be unsatisfactory). Nowadays, processing problems have been solved and CG cast irons are valuable because of their intermediate mechanical properties between gray and ductile iron, i.e an optimal combination of ductility and thermal conductivity (30-50 W/m.K), high thermal cyclic resistance and low weight that make CG cast irons specially useful for car parts fabrication [1,2].

Austempering heat treatments are widely used to improve ductility because they allow to produce a higher austenite fraction after cooling the material to room temperature [3]. In terms of properties, the Austempered Ductile Iron (ADI) matrix almost doubles the strength of conventional ductile iron while retaining its excellent toughness. Austempering does not produce the same type of structure in ductile iron as it does in steel because of the high carbon and silicon content of iron. The matrix structure of ADI as well as of austempered CG sets them apart from other cast irons, making them truly a separate class of engineering materials. Even though CG differs from ADI basically on the graphite morphology, it could be expected a close relation between CG and ADI microstructures, scarcely studied in the first case but extensively studied in the last one. The initial stage of austempering is the parent austenite transformation into ferrite plus high carbon austenite ($\gamma_{pa} \rightarrow \alpha_{Fe} + \gamma_{hc}$) at constant temperature, attaining a microstructure called bainite, i. e. fine laths of ferrite of approximately 0.2μm thick with interlath cementite of approximately 0.25μm [4]. It is well known for steels that bainite is produced at austempering temperatures in the range 523K-723K. Moreover, in steels with high enough Si content (~2wt%) carbide formation is inhibited because Si dissolves completely in ferrite and consequently austenite with high C content is expected to form in place of cementite [5]. The microstructure is then called bainitic ferrite [6]. The residual austenite that has left after transformation to bainite, exhibits two basic morphologies, *i.e.* film austenite which is retained between the subunits within a given sheaf of bainite and "blocky austenite" which is bounded by different bainite sheaves [7]. The distribution of carbon in those two forms of the residual austenite is not homogeneous after isothermal transformation to bainite. The austenite is enriched to a greater extent in the regions trapped between the platelets than in the blocky austenite [7]. The carbon concentration in austenite affects its chemical and mechanical stability at room temperature and the volume percent of retained austenite is vital to ductility and thoughness [8].

However, it must be noted that the microstructure obtained after austempering is not stable at room temperature, and thus the resulting material suffers an additional microstructural transformation during the final cooling. Thus, depending on the chemical composition, the material could be composed at room temperature of bainitic ferrite, retained austenite, graphite and even martensite.

It's important to notice that ADI mechanical properties are largely determined by its main constituents: bainitic ferrite and retained austenite [8]. Thus, it is important to go deeper in the understanding of *austenite → bainite* transformation to quantify the phases responsible of the final mechanical properties.

The transformation mechanism *austenite→ bainite* for cast irons is still under discussion and for CG cast irons was scarcely studied [9,10,11,12]. While some authors suggested a diffusion controlled phase transformation [9,13,14,15,16] some others found that a displacive mechanism controls the transformation in steels [5,17]. In the last case, it is thought that bainite subunits grow without diffusion, but that any excess carbon in the ferrite is partitioned into the residual austenite soon after nucleation [8,18]. Diffusion of interstitial carbon atoms would mostly operate during nucleation [19] redistributing carbon in a few milliseconds [9,10]; the kinetics is said to be controlled by the successive nucleation of laths or plates[6]. The indications are that the time required to grow a subunit is small relative to that needed to nucleate successive subunits. The growth rate of individual subunits is known to be much faster than the lengthening rate for sheaves [18]. Because growth occurs without diffusion, the transformation is said to be *interface controlled*.

On the other hand, according to the diffusive mechanism a short-range diffusion of the substitutional atoms is expected to occur at well-developed ledges in the austenite / ferrite interface [15,16]. It is also proposed that bainitic ferrite grows under full local equilibrium between ferrite and austenite what means that carbon content in ferrite would be prescribed by the $\alpha / \alpha + \gamma$ boundary in phase diagram [7,14]. In this case the transformation is said to be *diffusion controlled*.

The subject of this work is to progress in the understanding of the austempering process in vermicular cast irons. In particular, the main target is to obtain evidence that allows us to distinguish between the two kinetic mechanisms – displacive/interface or diffusion controlled growth – that have been proposed. Being available in the literature contradictory experimental data and theoretical arguments supporting both descriptions, it is clear that a global approach is necessary. It is important to have in mind that the proposed mechanism has to be consistent with both the microscopic and the macroscopic experimental evidence.

In this work, several microscopic experimental techniques were used for phase identification and microstructure characterization. Phases were quantified by X Ray Diffraction (XRD) patterns Rietveld analysis. Austenite fractions were determined by Transmission Mössbauer Spectroscopy (TMS) while Scanning Electron Microscopy (SEM) together with image analysis were used for acicular structures quantification. These structures were found to be composed by ferrite and austenite according to Transmission Electron Microscopy (TEM).

Subsequent modelling of the transformation was performed. First, an assessment of the free energy change involved in the *austenite$\rightarrow$ bainite* transformation was carried out assuming a plate bainite morphology. The surface free energy term was found to be negligible compared with the strain term and the most favourable plate length was of the order of the experimentally observed. Next, the kinetics of the transformation was modeled in the Avrami framework and compared to macroscopic experimental data. Both the diffusion controlled and the diffusionless growth hypotheses were considered in order to elucidate the mechanism underlying the austempering phase transformation.

Complementary experimental techniques, numerical simulations, image analysis, size distribution measurements and theoretical calculations are combined in this work for the first time to characterize vermicular cast irons microstructure. This combination of different approaches allows us to give a better description of the mechanisms that drive *austenite→ bainite* phase transformation which was never studied in the particular case of vermicular graphite morphology. Comparison between modelling and the available experimental data supports the hypothesis that the transformation is of displacive type in the analyzed material.

**EXPERIMENTAL**

Samples were prepared following the ASTMA-395 standard in a medium frequency furnace as quoted in references[10-12]. Sample composition was determined using chemical methods (Table 1). The heat treatment consisted of 30min at 1173 K and then quenching in a salt bath at 648 K for times between 1 min and 10 min. Finally, sample was air cooled down to room temperature (Figure 1).

**Table 1 Chemical composition in wt.% of the alloyed compacted graphite cast iron.**

| C | Si | Mn | Cu | P | S | Fe |
|---|---|---|---|---|---|---|
| 3.52 | 2.10 | 0.11 | 0.03 | 0.01 | 0.03 | 94.2 |

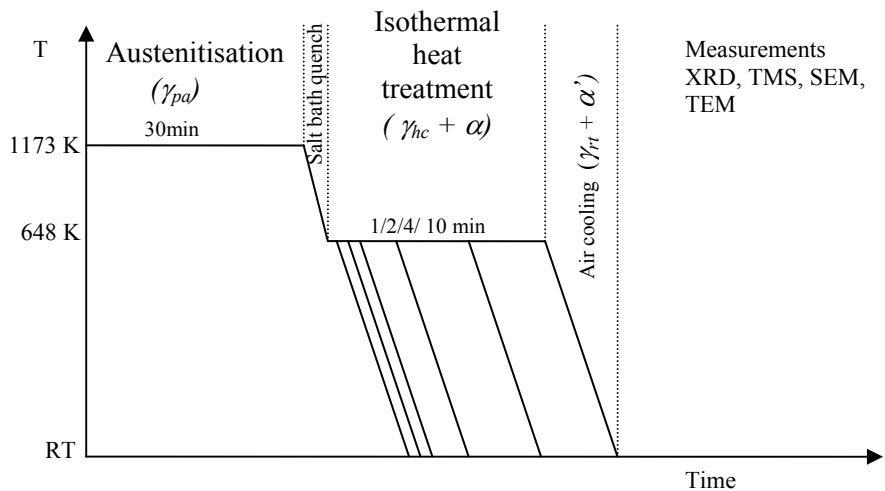

**Figure 1 Samples heat treatment. Italics will be used as phases nomenclature in the modelling section: $\gamma_{pa}$ ( ParentAustenite ), $\gamma_{hc}$ (High Carbon Thermal Austenite), $\alpha$ (Ferrite), $\gamma_{rt}$ (Retained Austenite at Room Temperature), $\alpha'$ (Martensite).**

Samples thicknesses were reduced to 70μm by conventional grinding techniques using diamond paste for posterior analysis by TMS[10-12].

In order to characterize the austempering microstructure by SEM, samples were etched with Nital 2% vol. For the case of TEM measurements, only the sample austempered for 2min was prepared using ion milling with two argon-ion guns.

Mössbauer spectra were taken in transmission geometry and spectrometer settings are described elsewhere[10-12].

X Ray diffraction patterns of a set of samples austemperized at a slightly different temperature (623K), prepared exactly in the same way, were taken in Bragg-Brentano geometry with a step mode collection as described in a previous work[20]. All measurements were done at room temperature in the angular range (40º-90º) with 10s/step. The Rietveld method was applied using the Full Prof program[21] and the fits were performed using austenite (Fm3m), ferrite (Im3m) and martensite (I4/mmm). To detect carbon graphite all samples were quickly scanned in a broader angular range (25º-125º). Graphite (P63mc) was added only for the 10min austempered sample Rietveld analysis. Goodness Rietveld fitting parameters were Rwp/Rexp =2, and 5% for each phase.

Quantification of structures detected by SEM was made with standard image processing software.

**RESULTS**

According to SEM images (Figure 2), three different regions can be recognized, i.e. dark vermicular particles, presumably carbon graphite, grey acicular structures and a light zone comprising the rest of the image. After 10 minutes of austempering treatment the quantity of acicular structures increases (Figure 2 b).

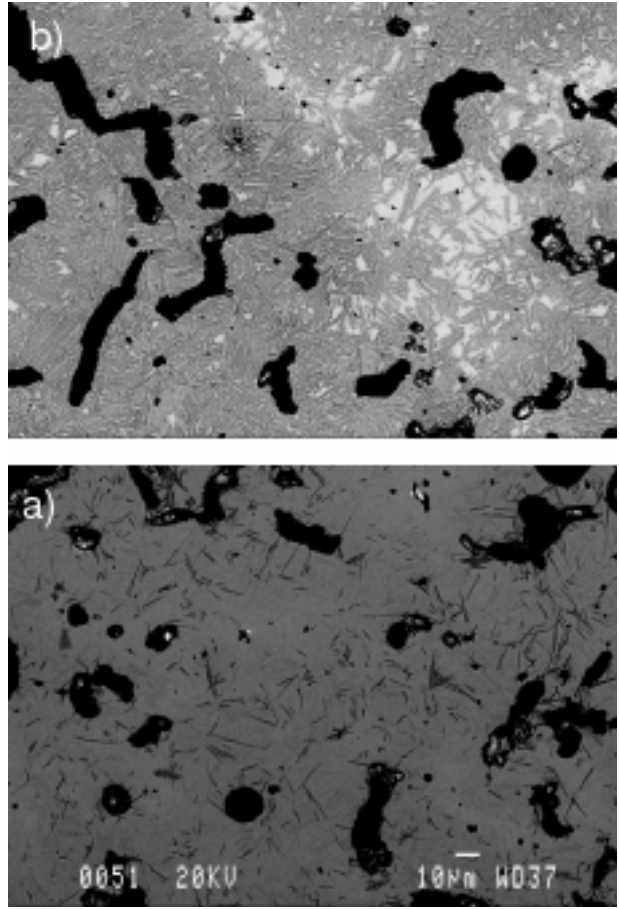

**Figure 2 SEM images after cooling to room temperature a sample austempered at 648K during a) 1 min b) 10 min.**

For the case of steels, specially when the alloy concentration is low, an expression was reported to calculate the initial temperature of martensitic transformation ($M_s$)[22]:

$$M_s (K) = 834 - 474 \times (wt\%C) - 33 \times (wt\%Mn) \quad (1)$$

In our case, the matrix carbon content while austenitizing can be estimated according to the expression [23]:

$$C^0_\gamma = -0.435 + 0.335 \times 10^{-3} \times T_\gamma + 1.61 \times 10^{-6} \times T_\gamma^2 + 0.006 \times (wt\%Mn) - 0.11 \times (wt\%Si)$$

where $C^0_\gamma$ is the austenite carbon content in wt% at the austenitizing temperature (1173K) and $T_\gamma$ is the austenitizing temperature in celsius degrees (900ºC). Thus according to this expression and samples composition we get $C^0_\gamma = 0.94$ wt%C.

Replacing $C^0_\gamma = 0.94$ wt%C in equation (1) $M_s(K)=385K$, well above room temperature. Even though this equation was proved to be useful for steels, we can consider the obtained value $M_s$ as an indicator of the posible existence of martensite at room temperature in our samples. In other words, martensite could be stable at room temperature if its carbon content is superior to 1,5 wt%C, which could be clearly the case in the samples analyzed here.

Those hypothesis were verified by XRD patterns where ferrite, austenite, martensite and carbon graphite were present in all samples (Figure 3).

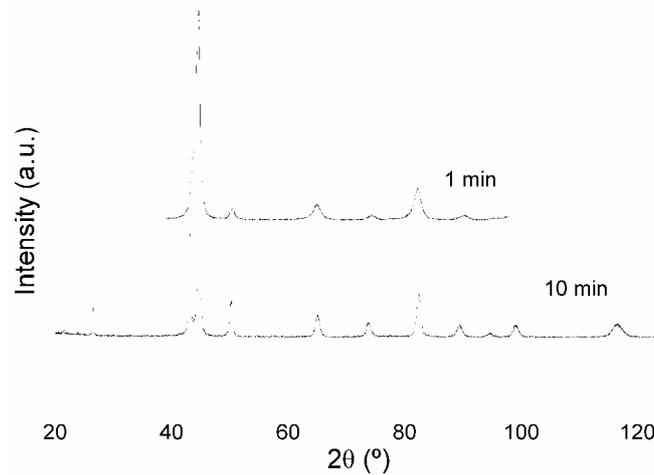

**Figure 3 Room temperature X ray diffraction patterns corresponding to samples austempered at 623K for 1minute and 10 minutes. The lines under the spectra correspond to the theoretical angles where, from top to bottom, ferrite, austenite, martensite and carbon graphite peaks appear.**

The total austenite fraction quantified by XRD and TMS is shown in Figure 4. Lattice constants were determined by the Rietveld method ($a_\alpha=2,89_1$ Å, $a_\gamma=3,654_9$ Å). The agreement is excellent even tough, as previously stated, XRD patterns were taken over a set of samples austemperized at a slightly different temperature (623K). It is important to note that TMS did not allow us to distinguish ferrite from martensite, because Fe local environments in martensite are similar to Fe sourroudings in ferrite[24]. Hence only austenite and ferrite plus martensite fraction was accurately quantified using TMS.

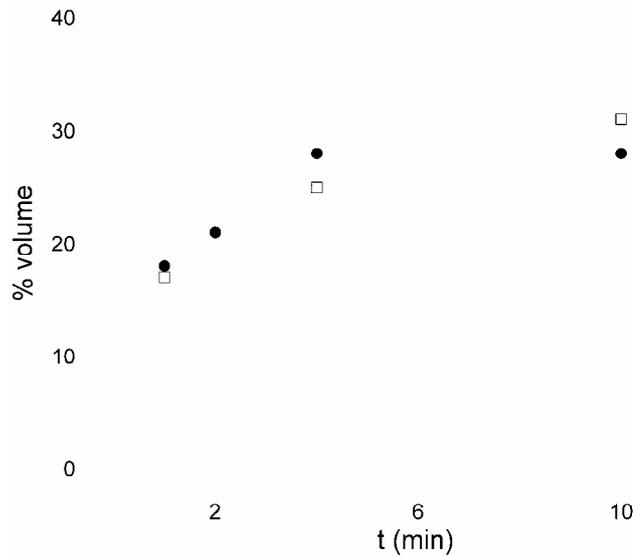
**Figure 4** Austenite fraction quantified by Ritveld analisys of XRD diffraction patterns for samples austempered at 623K (squares) and by TMS over samples austempered at 648K (circles) as a function of austempering time.

Quantification of acicular structures was performed through image analysis of SEM pictures, and the results are shown in Figure 5 as a function of the austempering time, as well as the amount of ferrite plus austenite assessed by XRD.

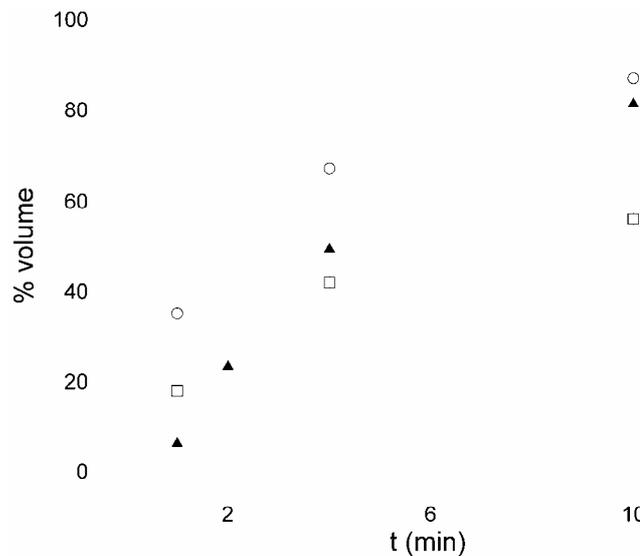
**Figure 5** Plot of several microstructural variables vs. austempering time for samples austempered at 648K. Surface fraction of acicular structures determined from SEM images (Triangles), fraction of ferrite (Squares) and fraction of ferrite + austenite (Circles), obtained by XRD for samples austempered at 623K.

According to Figure 5 the surface fraction of acicular structures (which is equivalent to the volume fraction for a random distribution of particles[25], falls between the ferrite and the ferrite plus austenite fractions after the first minute of austempering time. Taking into account some reported results for Austempered Ductile Iron (ADI)[26], austempering microstructure could be composed by ferrite laths with thin interlath films of austenite,

i. e. bainite. The bainitic transformation starts at a threshold temperature $B_s$ that for the case of steels has been empirically calculated taking into account the alloy composition[22]:

$$B_s(K) = 1103 - 270 \times (wt\%C) - 90 \times (wt\%Mn) - 37 \times (wt\%Ni) - 70 \times (wt\%Cr) - 83 \times (wt\%Mo)$$

Replacing $C^0_\gamma = 0.94$ wt%C and alloy composition (Table 1) in $B_s$ equation we obtained $B_s = 825$K which indicates that bainite could be expected in the samples analyzed here.

To check the presence of bainite, transmission electron microscopy (TEM) of one sample austempered for 2 minutes was performed. Electron diffraction patterns were obtained in several regions of the sample and the cell parameters for the ferrite and the austenite were calculated ($a_\alpha = 2.892 \pm 0.014$ Å, $a_\gamma = 3.654 \pm 0.009$ Å), in close agreement with the XRD results. These parameters are in accordance with the JCPDS files number 06-0696 for ferrite and 31-0619 for austenite, with relative errors of 0.9% and 0.2% respectively. The electron diffraction pattern confirms the presence of bainitic ferrite, i.e. austenite plus ferrite inside the acicular structures (Figure 6 and Figure 7). The size of the bainite sheave determined from TEM images is 4 μm long and 0.5 μm width, which is of the same order of magnitude than the reported for ADI and steels bainite sheaves [4, 5].

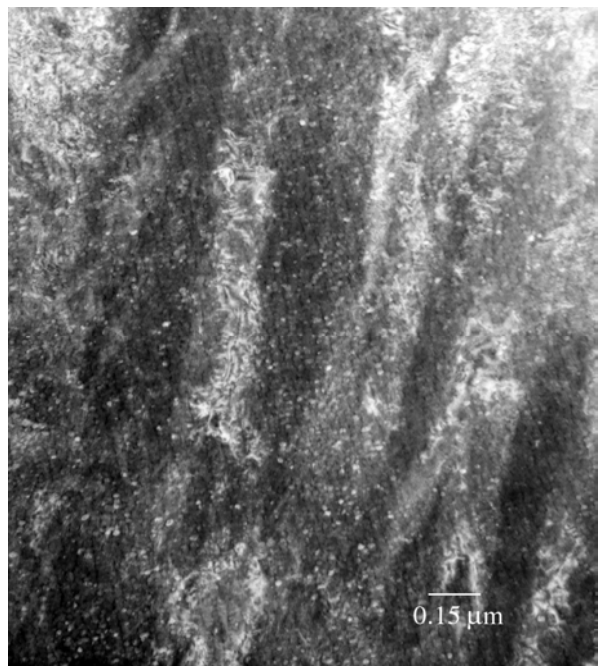

**Figure 6 TEM image of an acicular structure.**

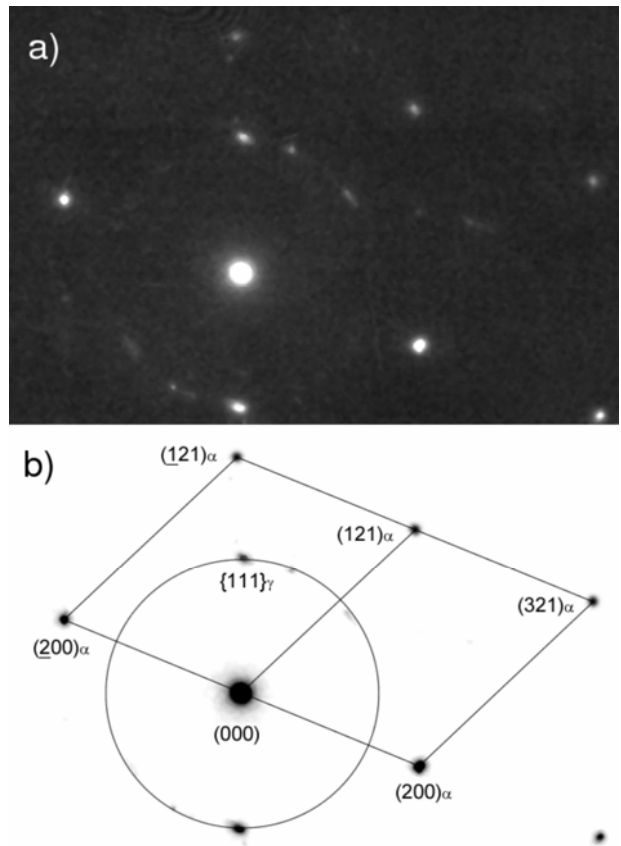

**Figure 7 (a) Electron diffraction pattern of a region in Figure 6. (b) Reciprocal space representation of (a).**

Thus, the acicular structures observed by SEM are identified as bainite sheaves, the 2-dimensional cut of the 3-dimension bainite plates.

As the ferrite plus austenite fraction obtained by XRD and shown in Figure 5 is greater than the fraction of bainitic acicular structures, that contain ferrite and austenite, it can be concluded that some austenite is stabilized during the final air cooling process and hence appear as retained austenite at room temperature. This result is consistent with reported observations for high silicon steels where residual austenite after transformation to bainite exhibits two basic morphologies, *i.e.* film austenite which is retained between the subunits within a given sheaf and blocky austenite which is bounded by different bainite sheaves [7]. Moreover, recent observations in-situ during austempering showed carbon-rich and carbon-poor regions in austenite [27]. Presumably austenite is enriched to a greater extent in the regions trapped between the platelets than in the blocky austenite [7].

Statistical distributions of the bainite sheave lengths are shown in Figure 8 for different austempering times. Statistics was made over approximately 1000 sheaves for each austempering time from SEM micrographs.

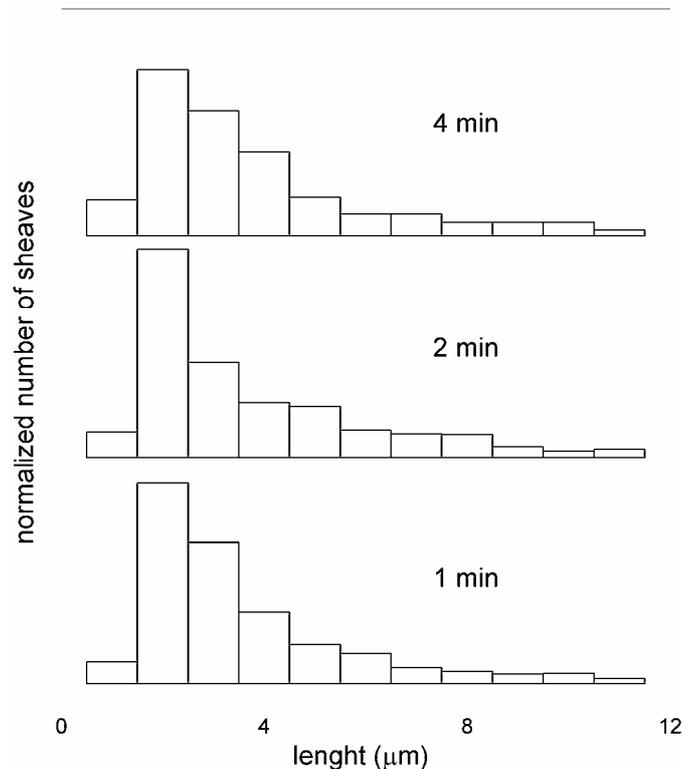

**Figure 8 Statistical distribution of bainite sheave lengths for samples austempered 1, 2, 4 minutes.**

Summarizing, the three regions identified in SEM images (Figure 2) are characterized as bainite (grey acicular structures), austenite plus martensite (the light grey background) and carbon graphite (vermicular black regions). Statistical length measurements shown in Figure 8 indicate that the length moda doesn't change with austempering time taking into account standard deviation, contrary to other authors conclusions over similar samples [9]. The fact that neither the shape nor the statistical moda of lengths distributions change with austempering time while the total number of sheaves increases, could indicate that length growth of plates is extremely slower than nucleation velocity. This result agrees with a recent in-situ observation during isothermal holding of high silicon steels, according to which bainitic widening/thickening and lengthening are formed by a shear mechanism and carbon diffusion appears to play a significant role previously to nucleation of the Bainite sheaves but not during its growth [28].

**Free energy assessment**

The free energy difference involved in the formation of a bainite nucleus of a given size within an austenite matrix can be written as [29]:

$$\Delta G = n\,(g^b - g^\gamma) + n\,\Delta g_s + \eta\,\sigma\,n^{\frac{2}{3}} \qquad (2)$$

accounting for the volume, strain and surface free energy contributions. Here, $g^b$ and $g^\gamma$ are the free energy per atom in the bainite and austenite phases respectively, $n$ is the number of atoms in the nucleus of bainite, $\Delta g_s$ is the elastic energy per atom, $\eta$ is a shape factor and $\sigma$ the interfacial free energy.

In the present case the free energy difference between bainite and austenite at 500ºC could be approximated from the steel case with a close chemical composition[30] $\Delta G^0_m = (g^b - g^\gamma) \approx -996$ J/mol.

Taking into account the diffraction constants determined by XRD, the number of atoms per unit volume in a nucleus of ferrite as well as of austenite were approximated as: $v^\alpha = 8{,}06$ $Å^3 at^{-1}$ and $v^\gamma = 6{,}10$ $Å^3 at^{-1}$ respectively; where the austenite phase was considered to have 2wt% of carbon. In a rough approximation, the number of atoms per unit volume in bainite was calculated as the average between $v^\alpha$ and $v^\gamma$, i. e. $v^b = 7{,}08$ $Å^3 at^{-1}$. Hence the first term in equation (2) reads:

$$\Delta G^{(1)} = n\,(g^b - g^\gamma) = -1{,}32 \times 10^{-9}\,\frac{J}{\mu m^3} \qquad (R^2\,y)$$

Moreover, the elastic energy per atom can be written as[21]:

$$\Delta g_s = \frac{2}{3}\left[\mu^\gamma\,\frac{(v^b - v^\gamma)}{v^b}\right] E\!\left(\frac{y}{R}\right)$$

where $\mu^\gamma$ is the shear modulus of the austenite, $v$ the volume per atom and $E$ is a shape function. The shear modulus of austenite was taken as $\mu_\gamma = 7 \times 10^{10}$ $Nm^{-2}$ [31].

Specifically, for a disk shape ($y/R \ll 1$) the $E$ function can be estimated as[29]:

$$E\left(\frac{y}{R}\right) \approx \frac{3}{4}\pi\frac{y}{R}$$

where $y$ becomes the semi-thickness of the disc and $R$ is its radius.

The second term in equation (2) was then computed as:

$$\Delta G^{(2)} = n\ \Delta g_s = 5,01\times10^{-8}\ \frac{J}{\mu m^3}\ (R\ y^2)$$

According to[29] the shape factor $\eta$ of an ellipsoid of revolution of semi-axes R, R and $y$ can be written as:

$$\eta = \left(\frac{3}{4}\pi v^b \frac{y}{R}\right)^{\frac{2}{3}}\left[2 + \frac{y^2}{R^2\sqrt{1-y^2/R^2}}\ \ln\left(\frac{1+\sqrt{1-y^2/R^2}}{1-\sqrt{1-y^2/R^2}}\right)\right]$$

Finally with an $\alpha/\gamma$ interface energy per unit area $\sigma_{\alpha/\gamma} = 0,2\ Jm^{-2}$ [32], the third term in the equation (2) was approximated as:

$$\Delta G^{(3)} = \eta\sigma n^{\frac{2}{3}} = 1,2\times10^{-12}\ \frac{J}{\mu m^2}\left(y^{\frac{4}{3}}R^{\frac{2}{3}}\right)\left[2 + \frac{y^2}{R^2\sqrt{1-y^2/R^2}}\ \ln\left(\frac{1+\sqrt{1-y^2/R^2}}{1-\sqrt{1-y^2/R^2}}\right)\right]$$

The most favorable nucleation path will use the shape which gives a saddle point on the total free energy difference ΔG. Assuming average ellipsoid semiaxis value of $y$=0,25μm (as observed by TEM) we obtained the free energy graph shown in Figure 9 where the saddle point match an average plate lenght of 8μm, which is of the same order than the extracted by TEM.

Moreover, assuming average ellipsoids semiaxis values $y$=0,25μm and $R$=2μm and replacing in previous equations we get $\Delta G^{(1)} = 1,32 \times 10^{-9}\ J$, $\Delta G^{(2)} = 6,26 \times 10^{-9}\ J$, $\Delta G^{(3)} = 6,26 \times 10^{-13}\ J$.

Even though this is an approximated calculus, the order of magnitude of the radius is close to the experimentally found by TEM and the surface free energy term results negligible compared with the others two, .supporting the assumption of plate formation.

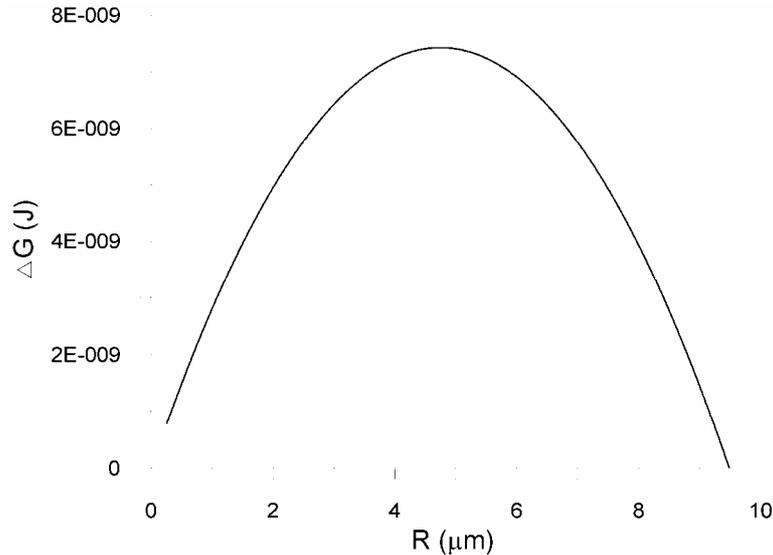

**Figure 9** Free energy plot of the transformation austenite → bainite for vermicular cast iron with chemical composition listed in Table 1, assuming a plate bainite morphology. The saddle point matches an average plate lenght of approximately 8μm.

## **Modelling**

The purpose of this section is to integrate all the collected information into a single description to distinguish between the two growth mechanisms proposed in the literature.

As a first assumption, we will assume that at the austenitizing temperature the samples are formed by a homogeneous matrix composed of austenite with carbon graphite inclusions. Being clearly separable inclusions, we will assume that carbon graphite has no influence in the austempering kinetics. Thus, at the austenizing temperature there is a single austenitic phase in equilibrium, and its volume fraction will be denoted by $x_{\gamma pa}$ (parent austenite), so initially $x_{\gamma pa}(0)=1$.

After quenching down to the austempering temperature, the austempering phase transformation *austenite → bainite* takes place. The $x_{\gamma pa}$ volume fraction decreases as the austempering time increases, an so a volume fraction (1- $x_{\gamma pa}$) proceeds through the *austenite* → *bainite* transformation. We will assume that during the austempering process a fraction *f* of the austenite transforms into ferrite and a fraction *(1-f)* transforms into high carbon, stable austenite (see Figure 1). Thus,

$$x_\alpha(t) = f \left[1 - x_{\gamma pa}(t)\right]$$

$$x_{\gamma hc}(t) = [1 - f] \left[1 - x_{\gamma pa}(t)\right]$$

where $x_\alpha$ and $x_{\gamma hc}$ are the ferrite and stable austenite volume fraction respectively, both phases composing bainite microstructure.

The evolution of phases with austempering time will be analysed in the frame of Avrami nucleation and growth theory[29]. In this description, the remaining $x_{\gamma pa}$ volume fraction is

$$x_{\gamma pa}(t) = e^{-x(t)}$$

being *x(t)* the extended volume of the $\alpha$ plus $\gamma_{hc}$ precipitates at time *t*.

After the austempering time $t_a$, the sample is cooled down to room temperature. The parent austenite ($x_{\gamma pa}$) becomes partially unstable, and we will assume that a fraction *g* transforms into martensite while the remaining fraction *(1-g)* becomes retained austenite at room temperature (see Figure 1). The corresponding relations are

$$x_{\alpha'} = g\, x_{\gamma pa}(t_a)$$

$$x_{\gamma rt} = [1 - g]\, x_{\gamma pa}(t_a)$$

where $\alpha'$ is the martensite volume fraction and $\gamma_{rt}$ is the retained austenite volume fraction that should be homogeneously distributed along the matrix.

According to this model the stable phases at room temperature are martensite (of volume fraction $x_{\alpha'}$), bainite (of volume fraction $x_\alpha + x_{\gamma hc}(t_a)$) and austenite (of volume fraction $x_{\gamma rt}$). Graphite fraction will be considered constant in time.

The actual behavior of these phases is defined by the chosen nucleation and growth model of the bainite phase. We will consider the two models proposed in the literature. Bainite plates will be modelled as ellipsoids of revolution, of axes *a* and *b* (revolution axis). According to microscopy observation and the free energy assessment of the previous section, critical values $a_0 = 0.5$ μm and $b_0 = 4$ μm are assumed.

*Difussion controlled growth*

Here we will consider that the growth of the bainite phase is controlled by the diffusion of C in the austenite phase. It is possible to considered that self-similar elliptic precipitates of axes a and b grow by diffusion. In a rough approximation:

$$a(t) = \sqrt{a_0^2 + Dt}$$

$$b(t) = \sqrt{b_0^2 + D\frac{b_0}{a_0}t}$$

where $D$ is the diffusion coefficient of C in austenite, that was determined to be $600\mu m^2 min^{-1}$ [33]. Denoting by $I$ the nucleation frequency, we obtain:

$$x(t) = I\int_0^t \tfrac{4}{3}\pi\sqrt{a_0^2 + D(t-\tau)}\left(b_0^2 + D\frac{b_0}{a_0}(t-\tau)\right)d\tau$$

Thus, the parameters to be determined in this model are the nucleation frequency I and the fractions f and g. The best fit parameters to the experimental data are given in Table 2 and the evolution of the different phases is shown in Figure 10.

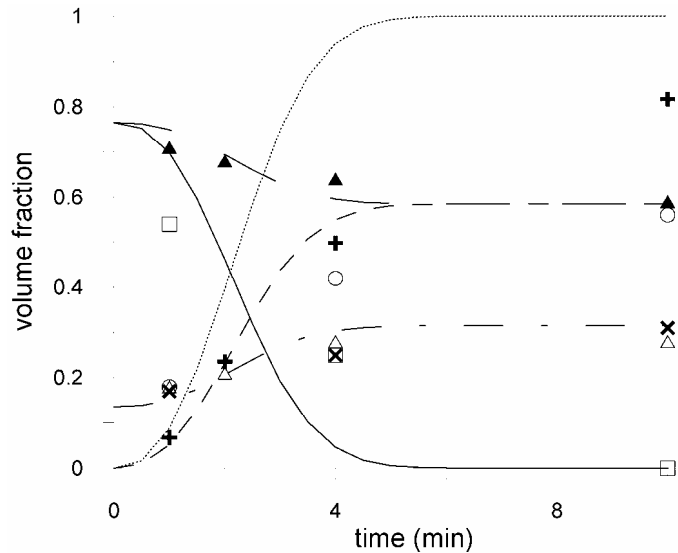

**Figure 10 Kinetics of the different phases assuming diffusion controlled growth (lines: − martensite, − · − austenite, − − ferrite, —— ferrite plus martensite and ······· bainite) compared to experimental data: austenite by XRD (×), ferrite by XRD (○), martensite by XRD (□), ferrite plus martensite (▲) and austenite (Δ) by Mössbauer spectroscopy and bainite by SEM (+).**

**Table 2 Best fit parameters obtained in the kinetics modelling.**

|  | f | g | I / $\mu m^{-3} min^{-1}$ | G / $\mu m\ min^{-1}$ |
|---|---|---|---|---|
| Diffusion controlled growth | 0.65 | 0.85 | $10^{-5}$ | - |
| Interface controlled growth | 0.65 | 0.85 | 0.0085 | 0.01 |

*Interface controlled growth*

We assume, as before, a growth rate in the ellipsoid axes a and b that maintains the self-similarity, i.e. the *a/b* quotient is constant through the time. We propose an Avrami-type kinetics with nucleation rate *I* and growth rate *G* constants in time. The expression can be written as:

$$x(t) = I\int_0^t \frac{4\pi}{3}(a_0 + G_a(t-\tau))(b_0 + G_b(t-\tau))^2 d\tau = \frac{4\pi}{3}I(At + Bt^2 + Ct^3 + Dt^4)$$

where $G_b=(b_0/a_0)G_a$, $G_a \equiv G$ and:

$$A = a_0^2 b_0$$

$$B = \frac{3}{2}G a_0 b_0$$

$$C = G^2 b_0$$

$$D = \frac{1}{4}G^3 \left(\frac{b_0}{a_0}\right)$$

Thus, this model is dependent on four parameters (*G, I, f* and *g*). The best fit parameters are given in Table 2, and the kinetics evolution of the different phases is shown in Figure11.

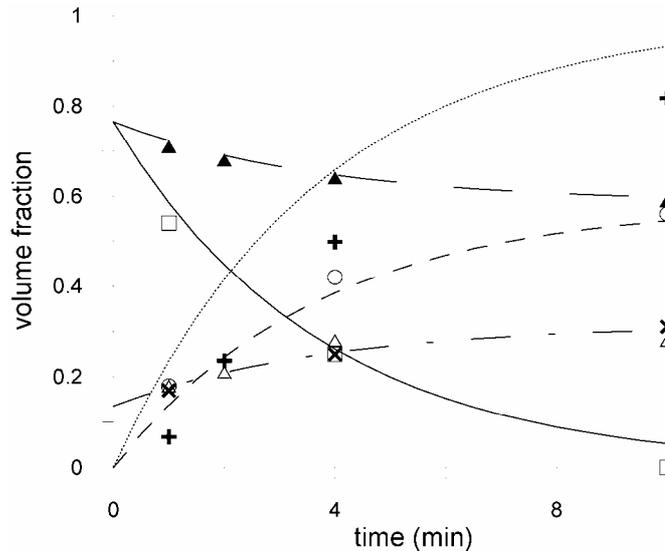

**Figure 11** Kinetics obtained in the interface controlled growth model (lines: − martensite, − · − austenite, − − ferrite, —— ferrite plus martensite and ········ bainite) compared to experimental data: austenite by XRD (×), ferrite by XRD (○), martenisite by XRD (□), ferrite plus martensite (▲) and austenite (Δ) by Mössbauer spectroscopy and bainite by SEM (+).

## **DISCUSSION AND CONCLUSIONS**

Both growth models give similar values of the fractions f and g that, on the other hand, cannot be determined experimentally. Thus, the validity of one model or the other must essentially rely on the plausibility of the values of the nucleation rate. SEM pictures for 1min austempering time were analyzed in order to count the number of acicular structures; from that measurement, and assuming an uniform distribution of particles within the sample volume, we can estimate the number of acicular structures as approximately $0.01 \mu m^{-3} min^{-1}$ which is the same order of magnitude of the nucleation frequency obtained in the best fit of the interface controlled growth model (Table 2). The observation of Figure10 and Figure11 seems also to indicate that the interface controlled growth kinetics is closer to the experimentally determined values. A deeper analysis of this kinetics shows that the value of the growth rate G is actually very low, which means that there is not significant growth of the bainite plates after nucleation. All this set of data is coherent with the displacive model proposed for the austenite →bainite transformation in steels[5].

Concerning to the sheave length distribution shown in Figure 8, in all cases a distinctive peak appears around 2 μm, slightly below the value of 4 μm used in the model. However, it must be considered that the two dimensional cut of the three dimensional bainite plates will always give a maximum at a value lower than the true radius of the plates due to stereologic reasons[25]. In these distributions a tail is also observed for lengths larger than that of a single plate, which is in well agreement with the fact pointed out by several authors that new bainite plates nucleate often close to the ends of existing plates[5,28].

To summarize, the austempering kinetics of a vermicular cast iron was analyzed by X-Ray Diffraction, Transmission Mössbauer Spectroscopy, Scanning and Transmission Electron Microscopy, and Image processing. Although every of the techniques gives a partial view of the transformation, the compound analysis allowed to offer a complete picture of the kinetics. The transformation was modeled in the framework of the Avrami kinetics, assuming that the austempering process produces the transformation of austenite in bainite and the final cooling down to room temperature induces the partial decomposition of the untransformed austenite (unstable at room temperature) in martensite and retained austenite (stable at room temperature). Two different growth mechanisms, namely diffusion controlled and interface controlled, were considered to model the transformation, and comparison with the collected experimental data shows a much better agreement with the interface controlled growth. Thus, it appears that the mechanism of the austempering process is very close to the displacive mechanism of the formation of bainite proposed by some authors[5,6,28,34], in which the diffusion of C is the responsible of the nucleation process of the bainite sheaves, that appear as a consequence of a localized displacive transformation when the C concentration is adequate, but further growth of the bainite plates is almost suppressed.

**Acknowledgements**

The authors want to acknowledge specially to Prof. Judith Desimoni and Dr. Ricardo Gregorutti for the collaboration with TMS, XRD measurements and samples preparation.

This work was funded by CONICET, Argentina, CICYT, grant MAT2004-01214, and Generalitat de Catalunya, grants SGR2005-00535 and SGR2005-00201.